\documentclass[prl,preprint,showpacs,floatfix]{revtex4-1}
\usepackage{psfrag}
\usepackage{amssymb,amsmath,amsthm}
\usepackage[dvips]{graphicx}
\usepackage{verbatim}
\usepackage{ulem}
\usepackage{cancel}
\usepackage[colorlinks=true,linkcolor=blue,citecolor=blue]{hyperref}
\renewcommand\Re{\operatorname{Re}}
\renewcommand\Im{\operatorname{Im}}

\begin{document}

\title{Delayed feedback control of self-mobile cavity solitons}
\date{\today}
\author{A. Pimenov$^{1}$, A. G. Vladimirov$^{1,2}$, S. V. Gurevich$^{3}$, K. Panajotov$^{4,5}$, G. Huyet$^{2}$, and M. Tlidi$^{6}$}
\affiliation{$^{1}$Weierstrass Institute, Mohrenstrasse 39, D-10117 Berlin, Germany}
\affiliation{$^{2}$ Cork Institute of Technology, Rossa Ave, Bishopstown, Cork, Ireland}
\affiliation{$^{3}$Institute for Theoretical Physics, University of M{\"u}nster, Wilhelm-Klemm-Str.9, D-48149, M{\"{u}}nster, Germany}
\affiliation{$^{4}$Department of Applied Physics and Photonics (IR-TONA), Vrije Unversiteit Brussels, Pleinlaan 2, B-1050 Brussels, Belgium}
\affiliation{$^{5}$Institute of Solid State Physics, 72 Tzarigradsko Chaussee Blvd., 1784 Sofia, Bulgaria}
\affiliation{$^{6}$Universit\'{e} Libre de Bruxelles (U.L.B.), Facult{\'{e}} des Sciences,  CP. 231, Campus Plaine, B-1050 Bruxelles, Belgium}

\begin{abstract}
Control of the motion of cavity solitons is one the central problems in nonlinear optical pattern formation. We report on the impact of the phase of the time-delayed optical feedback  and carrier lifetime on the self-mobility of localized structures of light in broad area semiconductor cavities. We show both analytically and numerically that the feedback phase strongly affects the drift instability threshold as well as the velocity of cavity soliton motion above this threshold. In addition we demonstrate that non-instantaneous carrier response in the semiconductor medium is responsible for the increase in critical feedback rate corresponding to the drift instability. 
\end{abstract}

\maketitle

The emergence of spatial-temporal dissipative structures far from equilibrium is a well documented issue since the seminal works of  Turing \cite{Turing}, Prigogine, and Lefever \cite{Prigogine1}. Dissipative structures have been theoretically predicted and experimentally observed in numerous nonlinear chemical, biological, hydrodynamical, and optical systems (for reviews on this issue see \cite{Review1,Review2}). They can be periodic or localized in space \cite{Barland}. Localized structures of light in nonlinear laser systems often called cavity solitons are among the most interesting  spatiotemporal patterns  occurring in extended nonlinear systems. They have attracted growing interest in optics due to potential applications for all-optical control of light, optical storage, and information  processing \cite{Barland}. 

Localized structures may loose their stability and start to move spontaneously as a result of symmetry breaking drift bifurcation due to finite relaxation time \cite{TSB93}-\cite{Gurevich} or  delayed feedback \cite{Tlidi1}-\cite{Gurevich13}. The motion of cavity soliton can also be triggered by an external  symmetry breaking effects such as a phase gradient \cite{Turaev08}, a symmetry breaking due to off-axis feedback \cite{Ramazza2}, or resonator detuning \cite{Kestas}, and an Ising-Bloch transition \cite{Coullet,Michaelis,Staliunas}. 
\begin{figure}[tbp]
\begin{center}
 \includegraphics[width=0.8\linewidth]{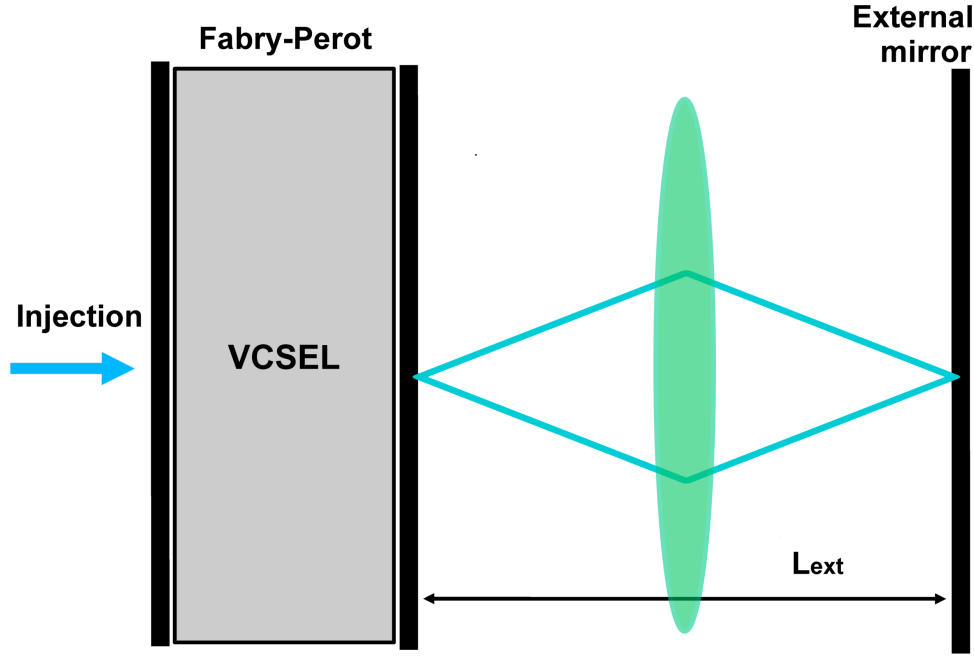}
\end{center}
\caption{Schematic setup of a nonlinear Fabry-Perot cavity  based on a vertical-cavity surface emitting laser (VCSEL) structure, driven by a coherent externally injected beam. The cavity is subject to delayed self-imaging optical feedback from an external mirror located at a distance $L_{ext}$ from the VCSEL output facet.}
\label{fig:setup}
\end{figure}

In what follows we investigate a drift instability of the cavity solitons induced by the time-delayed feedback, which provides a robust and a controllable mechanism, responsible for the appearance of a spontaneous motion. Moving localized strutures and fronts were predicted to appear in extended nonlinear optical ~\cite{Tlidi1}-\cite{Gurevich13} and population dynamics \cite{ecology} systems,  as well as in several chemical and biological systems, described by reaction-diffusion models~\cite{reaction-diffusion}. Previous works revealed that when the product of the delay time and the rate of the feedback exceeds some threshold value,  cavity solitons start to move in an arbitrary direction  in the transverse plane \cite{Tlidi1}-\cite{Gurevich13}. In these studies, the analysis was restricted to the specific case of nascent optical bistability described by the real Swift-Hohenberg equation with a real feedback term.   

The purpose of the present Letter is to study the role of the phase of the delayed feedback and the carrier lifetime on the motion of cavity solitons in broad-area semiconductor cavities. This simple and robust device received special attention owing to advances in semiconductor technology. We show that  for certain values of the feedback phase cavity soliton can be destabilized via a drift bifurcation leading to a spontaneous motion in the transverse direction. Furthermore, we demonstrate that the slower is the carrier decay rate in the semiconductor medium, the higher is the threshold associated with the motion of cavity solitons. Our analysis has obviously a much broader scope than semiconductor cavities and could be applicable to large variety of optical and other systems.  

We consider a broad-area semiconductor cavity operating below the lasing threshold and  subject  to a coherent optical injection and an optical feedback from a distant mirror in a self-imaging configuration, see Fig.~\ref{fig:setup}. The time-delayed feedback  is modeled according to Rosanov-Lang-Kobayashi-Pyragas approach \cite{rosanov75}. This device can be described by the following dimensionless equations \cite{KM1,KM2}
\begin{eqnarray}
\frac{dE}{dt} &=& -\left(\mu+i\theta\right)E + 2C(1-i\alpha)(N-1)E  \label{eq:dEdt} \\
&+& E_{i} +\eta e^{i\varphi}[E(t)-E(t-\tau)]+ i\nabla^{2} E\,, \nonumber \\
\frac{dN}{dt} &=& -\gamma \left[ N -I + (N-1)\left|
E\right| ^{2} - d\nabla^{2} N\right]\,, \label{eq:dNdt}
\end{eqnarray}
where $E$ is the slowly varying  electric field envelope and $N$ is the carrier density. The parameter $\alpha$ describes the linewidth enhancement factor, $\mu={\tilde\mu+\eta\cos{\varphi}}$, and $\theta={\tilde\theta}+\eta\sin{\varphi}$, where ${\tilde\mu}$ and ${\tilde\theta}$ are the cavity decay rate and the cavity detuning parameter, respectively. Below we will assume $\eta$ to be small enough, so that we can neglect the dependence of the parameters $\mu$ and $\theta$ on $\phi$. The parameter $E_{i}$ is the amplitude of the injected field,  $C$ is the bistability parameter, $\gamma$ is the carrier decay rate, $I$ is the injection current, and $d$ is the carrier diffusion coefficient. The diffraction of light and the diffusion of the carrier density are described by the terms $i\nabla^{2}E$ and $d\nabla^{2}N$, respectively, where $\nabla^{2}$ is the Laplace operator acting in the transverse plane $(x,\,y)$. The feedback is characterized by the delay time $\tau=2L_{ext}/c$, the feedback rate $\eta\ge 0$, and phase $\varphi$, where $L_{ext}$ is the external cavity length, and $c$ is the speed of light. 
\begin{figure}[tbp]
\begin{center}
\includegraphics[width=9.cm]{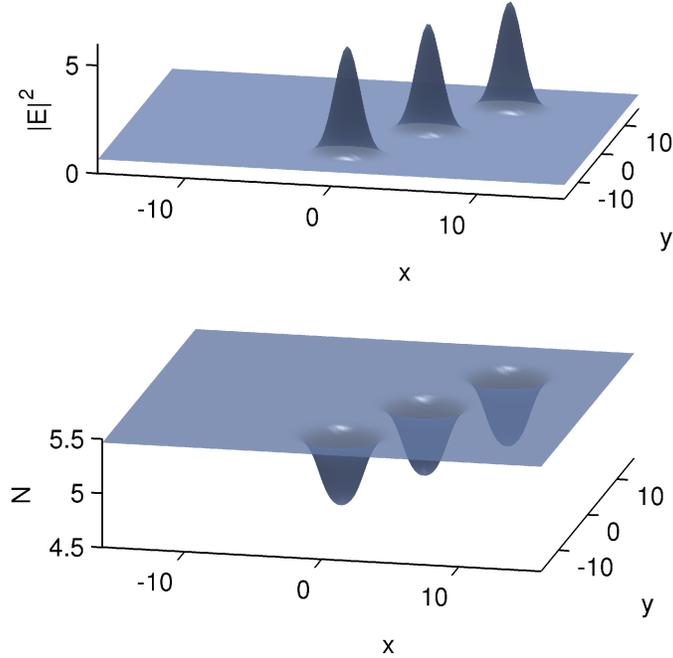}
\end{center}
\caption{Field intensity (top) and carrier density $N$ (bottom) of two-dimensional moving cavity soliton at different times. The laser parameters are $C = 0.45$, $\theta = -2$, $\alpha = 5$, $\gamma = 0.05$, $d = 0.052$, $\mu = 2$. The injection is $E_i=0.8$, and the feedback parameters are $\tau = 200$, $\eta = 0.07$, and $\varphi = 3.5$. Split-step Fourier method was used to obtain solitons at $t = 28000$, $t=30000$, and $t=32000$ (from right to left).}\label{fig:2}
\end{figure}

In the absence of delayed feedback, $\eta=0$,  we recover the mean field model \cite{Spinelli_pra98}, which supports stable stationary patterns and localized structures \cite{SC-semiC}. In the case of one spatial dimension the localized solutions correspond to homoclinic solutions of Eqs.~(\ref{eq:dEdt}) and (\ref{eq:dNdt}) with $\frac{dE}{dt}=\frac{dN}{dt}=0$ and the transverse coordinate $x$ considered as a time variable. They are generated in the subcritical domain, where a uniform intensity background and a branch of spatially periodic pattern are both linearly stable \cite{Barland}. When the feedback rate $\eta$ exceeds a certain critical value, a cavity soliton can start to move, see Fig.~\ref{fig:2}. Since the system is isotropic in the $(x,y)$-transverse plane, the velocity of the moving soliton has an arbitrary direction.
Numerical simulations were performed using the split-step Fourier method with periodic boundary conditions in both transverse directions.

To calculate the critical value of the feedback rate, which corresponds to the drift instability threshold, and small cavity soliton velocity $v=|{\mathbf v}|$ near this threshold, we look for a solution of Eqs.~(\ref{eq:dEdt}) and (\ref{eq:dNdt}) in the form of a slowly moving cavity soliton expanded in power series of $v$: $E = E_0(\boldsymbol{\xi}) + v \, \left[E_1(\boldsymbol{\xi}) + v E_2(\boldsymbol{\xi})+v^2 E_3(\boldsymbol{\xi})+...\right]$ and $N = N_0(\boldsymbol{\xi}) + v\,\left[N_1(\boldsymbol{\xi})+v N_2(\boldsymbol{\xi}) + v^2 N_3(\boldsymbol{\xi})+...\right]$. Here $E=E_0(\mathbf{r})$ and $N=N_0(\mathbf{r})$ is the stationary soliton profile, $\boldsymbol{\xi}=\mathbf{r}-v\,\mathbf{e}\,t$, $\mathbf{r}=(x,\,y)$, and $\mathbf{e}$ is the unit vector in the direction of the soliton motion. Substituting this expansion into Eqs.~(1) and (2) and collecting the first order terms in small parameter $v$ we obtain:

\begin{eqnarray}
L \left(\begin{array}{l}\Re E_{1}\\ \Im E_{1}\\N_1\end{array}\right) = \left(\begin{array}{l}\Re(w_{01} - \eta \tau w_{01} e^{i\varphi})\\ \Im(w_{01} - \eta \tau w_{01} e^{i\varphi}) \\\gamma^{-1}m_{01}\end{array}\right)\label{eq:1storder}
\end{eqnarray}
with $w_{01}={\mathbf e}\cdot \nabla E_0$ and $m_{01}={\mathbf e}\cdot \nabla N_0$.
The linear operator $L$ is given by
$$
L = \left(\begin{array}{ccc}\mu - 2 Cn_0 & \nabla_{eff}^2 & -2C(A_{0} + \alpha B_{0})\\
-\nabla_{eff}^2  &\mu-2Cn_0 & -2 C (B_{0} - \alpha A_{0})\\
 2n_0 A_{0} & 2n_0 B_{0} &-d \nabla^2+1+|E_0|^2 \end{array}\right),
$$
where $A_0=\Re E_0$, $B_0=\Im E_0$, $\nabla_{eff}^2=\nabla^2-\theta- 2 C\,\alpha\, n_0$, and $n_0=N_0-1$. By applying the solvability condition to the right hand side of Eq.~(3), we obtain the drift instability threshold
\begin{equation}
\eta =\eta_0=\frac{1}{\tau}\frac{1+\gamma^{-1}b }{ \sqrt{1+a^2}\cos(\varphi+\arctan{a})}
\label{thresholdlimit}
\end{equation}
with $a=(\langle \Re w_{01}, Y\rangle - \langle \Im w_{01}, X \rangle)/g$, $b=\langle m_{01}, Z \rangle/g$, and $g=\langle \Re w_{01},\, X\rangle + \langle \Im w_{01},\,Y \rangle$. Here, the eigenfunction ${\psi}^{\dagger}=\left(X,\,Y,\,Z\right)^T$  is the solution of the homogeneous adjoint problem $L^\dagger \psi^{\dagger} = 0$ and the scalar product $\langle \cdot \rangle$ is defined as  $\langle \psi_1,\, \psi_2\rangle= \int_{-\infty}^{+\infty}\psi_{1}\psi_{2}^\dagger\,d{\mathbf r}$. To estimate the coefficients $a$ and $b$ we have calculated the function $\psi$ numerically using the relaxation method in two transverse dimensions, $(x,y)$. 

It is noteworthy that since the stationary soliton solution does not depend on the carrier relaxation rate $\gamma$, the coefficients $a$ and $b$ in the threshold condition (\ref{thresholdlimit}) are also independent of $\gamma$.
In the limit of instantaneous carrier response, $\gamma \rightarrow \infty$, and zero feedback phase, $\varphi=0$, we recover from (\ref{thresholdlimit}) the threshold condition obtained earlier for the drift instability of cavity solitons in the Swift-Hohenberg equation with delayed feedback, $\eta_0 = \tau^{-1}$\cite{Tlidi1}. Note that the drift instability exists in Eqs.~(\ref{eq:dEdt}) and (\ref{eq:dNdt}) only for those feedback phases when the cosine function is positive in the denominator of Eq.~(\ref{thresholdlimit}). Furthermore, at $\gamma\to\infty$, $a\neq 0$, and $\varphi=-\arctan{a}$ the critical feedback rate appears to be smaller than that obtained for the real Swift-Hohenberg equation, $\eta_0=\tau^{-1}/\sqrt{1+a^2}<\tau^{-1}$. 

In order to calculate the first order corrections $E_1$ and $N_1$ to the stationary soliton solution $E_0$ and $N_0$ we have solved the system (\ref{eq:1storder}) numerically using the relaxation method. The second order corrections $E_2$ and $N_2$ have been obtained in a similar way by equating the second order terms in the small parameter $v$. Finally, assuming that a small deviation of the feedback rate from the drift bifurcation point (\ref{thresholdlimit}) is of the order $v^2$ and equating the third order terms in $v$, we obtain  
\begin{figure}
\begin{center}
\includegraphics[width=8.cm]{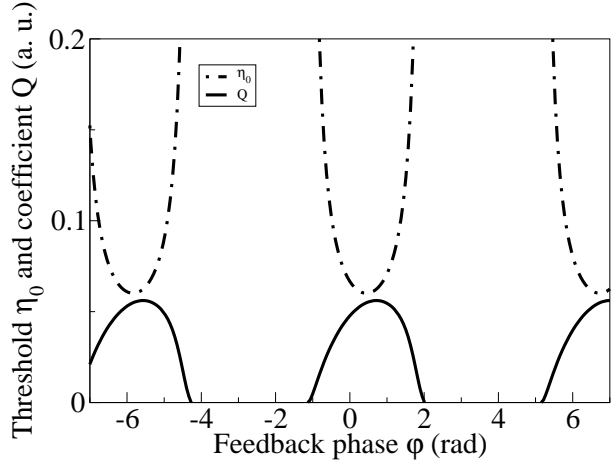}
\end{center}
\caption{ Threshold $\eta_0$ associated with the moving cavity soliton (dash-dotted line) given by Eq.~(\ref{thresholdlimit}) and coefficient $Q$ in the expression for the soliton velocity~\eqref{eq:long} calculated numerically using Eq.~(\ref{eq:Q})  (solid line). The parameters are $\mu = 1$, $\theta=-2$, $C=0.45$, $\alpha=5$, $\gamma=0.05$, $\tau=200$, $d = 0.052$, $E_i = 0.8$, $I = 2$.}\label{fig:3}
\end{figure}

\begin{figure}
\begin{center}
\includegraphics[width=8.cm]{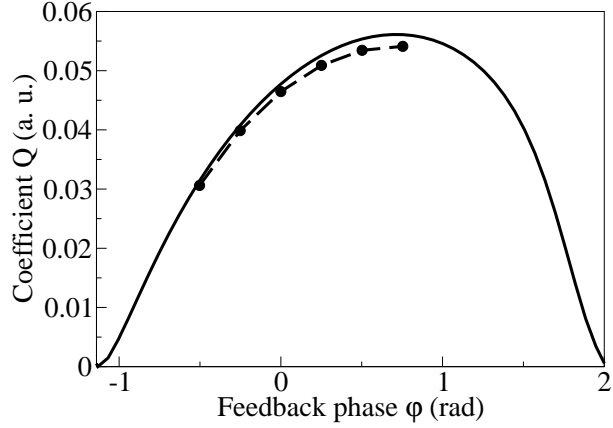}
\end{center}
\caption{Coefficient $Q$ in the expression for the soliton velocity~\eqref{eq:long} calculated numerically as a function of the phase $\varphi$ of the delayed feedback  (solid line). The circles indicate the values of $Q$ estimated by direct numerical integration of Eqs.~(\ref{eq:dEdt}) and (\ref{eq:dNdt}) using the split-step Fourier method. Parameters are the same as in Fig.~\ref{fig:3}.}\label{fig:4}
\end{figure}

\begin{figure}
\begin{center}
\includegraphics[width=8.cm]{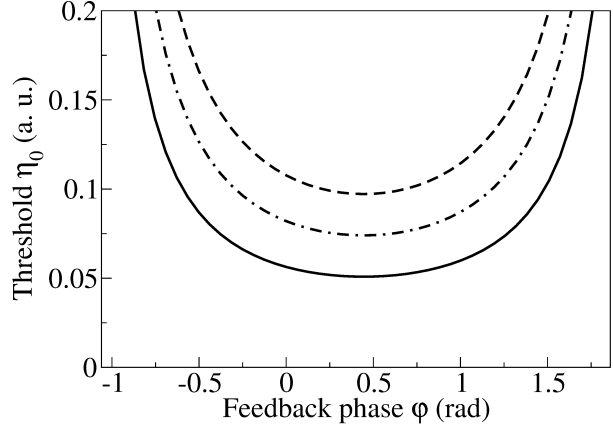}
\end{center}
\caption{Threshold associated with the motion of cavity soliton as a function of the phase of delayed feedback obtained for  different values of the carrier decay rate: $\gamma=0.06$ (solid), $\gamma=0.04$ (dash-dotted), $\gamma=0.03$ (dashed). Other parameters are the same as in Fig.~\ref{fig:3}.}\label{fig:5}
\end{figure}

\begin{equation}
v^2 L \begin{pmatrix}\Re E_3\\\Im E_3\\ N_3\end{pmatrix} = 
\begin{pmatrix}\Re [-r_1 + v^2 (\eta_0\tau e^{i\varphi}W_{\tau}+r_3)]\\\Im [-r_1 + v^2(\eta_0\tau e^{i\varphi}W_{\tau}+r_3)]\\v^2 (\gamma^{-1} m_{21}+ r_n)\end{pmatrix},\,\label{eq:trd1}
\end{equation}
\begin{eqnarray*}
 r_{1} &=&(\eta- \eta_0) e^{i \varphi} \tau w_{01},\\
 W_{\tau} &=& \tau^2 w_{03}/6 + \tau w_{12}/2 + w_{21}, \\
 r_{3} &=& w_{21}+2 C(1 - i \alpha) (N_1 E_2+N_2 E_1),\\
 r_n &=& -N_1 |E_1|^2- N_2 (E_0^* E_1 + E_0 E_1^*) - \\
&-&N_1 (E_0^* E_2 + E_0 E_2^*) - n_0 (E_1 E_2^*+E_2 E_1^*)\,,
\end{eqnarray*}
where $w_{03}={\mathbf e}\cdot\nabla({\mathbf e}\cdot\nabla({\mathbf e}\cdot\nabla E_0))$,  $w_{12}={\mathbf e}\cdot\nabla({\mathbf e}\cdot\nabla E_1)$, $w_{21}={\mathbf e}\cdot\nabla E_2$, and $m_{21}={\mathbf e}\cdot\nabla N_2$. The solvability condition for Eq.~(\ref{eq:trd1}) requires orthogonality of the right hand side of this equation to the eigenfunction $\psi^{\dagger}$ of the adjoint linear operator $L^{\dagger}$. This condition yields the following expression for the soliton velocity: 
\begin{eqnarray}
  v&=&  (\eta-\eta_0)^{1/2}Q,\label{eq:long}\\
  Q^2&=& \frac{\tau\, g\sqrt{1+a^2}\cos{(\varphi+\arctan{a})}}{\eta_0\,\tau h_{\tau}\sqrt{1+p_{\tau}^2}\cos{(\varphi+\arctan{p_{\tau}})}+q+s/\gamma}\,\label{eq:Q}
\end{eqnarray}
with $h_{\tau}=\langle \Re W_{\tau},X\rangle+\langle \Im W_{\tau},Y\rangle$, $p_{\tau}=(\langle \Re W_{\tau},Y\rangle- \langle \Im W_{\tau},X\rangle)/h$, $q=\langle r_{n},Z\rangle$, and $s=\langle m_{21},Z\rangle$.
Here, $h_{\tau}=\langle \Re W_{\tau},X\rangle+\langle \Im W_{\tau},Y\rangle$, $p_{\tau}=(\langle \Re W_{\tau},Y\rangle- \langle \Im W_{\tau},X\rangle)/h$, $q=\langle r_{n},Z\rangle$, and $s=\langle m_{21},Z\rangle$.
In the case of zero feedback phase and instantaneous carrier response, $\gamma^{-1}=\varphi=0$, the change of the soliton shape induced by its spontaneous motion is of order $v^3$ \cite{Gurevich13} ($E_{1}=N_{1}=E_{2}=N_{2}=0$), and, hence, we have $q=r_{n}=s=0$ and $W_{\tau}=\tau^2w_{03}/6$. Therefore, in this case the expression for the coefficient $Q$ in Eq.~(\ref{eq:long}) is transformed into $Q^2=g/(6\tau^2h')$  with $h'=\langle \Re w_{03},X\rangle+\langle \Im w_{03},Y\rangle$, and we recover a result similar to that obtained earlier for the real Swift-Hohenberg equation with delayed feedback \cite{Tlidi1}, $v^2=(\eta-\eta_0)g/(6\tau^2h')$.

The dependence of the critical feedback rate $\eta_0$ on the feedback phase $\varphi$, calculated using Eq.~(\ref{thresholdlimit}) is shown in Fig.~\ref{fig:3} by the dash-dotted curves.  One can see, that the drift instability exists only within the subinterval $(\varphi_{min}-\pi/2,\varphi_{min}+\pi/2)$ of the interval $(\varphi_{min}-\pi,\varphi_{min}+\pi)$, where $\varphi_{min}=-\arctan{a}$ is the feedback phase, corresponding to the lowest critical feedback rate $\eta_0^{min}=(1+\gamma^{-1}b)/(\tau\sqrt{1+a^2})$. In addition,  $\eta_0$ increases very rapidly when approaching the boundaries of this subinterval.  The solid lines in Fig.~\ref{fig:3} illustrate the dependence of the coefficient $Q$, defined by Eq.~(\ref{eq:Q}), on the phase $\varphi$. This coefficient determines the growth rate of the soliton velocity with the square root of the deviation of the feedback rate from the drift bifurcation point, $(\eta-\eta_0)^{1/2}$. Finally, it is seen from Fig.~\ref{fig:4} that the values of the  coefficient $Q$ obtained from Eq.~(\ref{eq:Q}) are in a good agreement with those of the quantity $v(\eta-\eta_0)^{-1/2}$ estimated by calculating the soliton velocity near the drift instability threshold with the help of direct numerical simulations of the model equations (\ref{eq:dEdt}) and (\ref{eq:dNdt}). 
%
%

The impact of carrier decay rate $\gamma$ on the soliton drift instability threshold is illustrated in Fig.~\ref{fig:5}. It is seen that the threshold value of the feedback rate $\eta_0$ increases with $\gamma$, which indicates that the coefficient $b$ in Eq.~\ref{thresholdlimit} must be positive. Thus, non-instantaneous carrier response in a semiconductor cavity leads to a suppression of the drift instability of cavity solitons.
%

To conclude, we have shown analytically and verified numerically that the mobility properties of transverse localized structures of light in a broad area semiconductor cavity with delayed feedback are strongly affected by the feedback phase. In particular, the drift instability leading to a spontaneous motion of cavity solitons in the transverse direction can develop with the increase of the feedback rate only in a certain interval of the feedback phases. Furthermore, we have demonstrated that the critical value of feedback rate corresponding to the drift instability threshold is higher in the case of a semiconductor cavity with slow carrier relaxation rate than in the instantaneous nonlinearity case. The results presented here constitute  a practical way of controlling the mobility properties of cavity solitons in broad area semiconductor cavities.

A.P. and A.G.V. acknowledge the support from SFB 787 of the DFG. A.G.V. and G.H. acknowledges the support of the EU FP7 ITN PROPHET and E.T.S.  Walton Visitors Award of the Science Foundation Ireland. With the support of the F.R.S.-FNRS. This research was supported in part by the Interuniversity Attraction Poles program of the Belgian Science Policy Office under Grant No. IAP P7-35.

\end{document}